\documentclass[adp,a4paper,fleqn]{w-art}
\usepackage{times,cite,w-thm}
\theoremstyle{plain}

\theoremstyle{definition}

\usepackage[]{graphicx}

\begin{document}

\DOIsuffix{theDOIsuffix}
\Volume{524}
\Month{02}
\Year{2012}

\pagespan{1}{}

\Receiveddate{1 December 2011}
%\Reviseddate{XXXX}
%\Accepteddate{XXXX}
%\Dateposted{XXXX}

\title[]{EXPERT OPINION \\ \\
         The apparent Fermi liquid concept helps to understand \\
         thermoelectric materials}

\author[]{Andrzej M. Ole\'s
          \footnote{E-mail:~\textsf{a.m.oles@fkf.mpg.de}}}

\address[]{Max-Planck-Institut f\"ur Festk\"orperforschung,
           Heisenbergstrasse 1, D-70569 Stuttgart, Germany}
\address[]{Marian Smoluchowski Institute of Physics, Jagellonian
           University, Reymonta 4, PL-30059 Krak\'ow, Poland}

\maketitle

There has been considerable recent interest in novel thermoelectric
materials suitable for applications such as waste heat recovery. In
general these materials are in the regime where electron-phonon
interaction or spin fluctuations are important as opposed to low
temperature, where point defect and electron-electron scattering
usually dominate and the Sommerfeld expansion applies. The thermopower
$S(T)$ and its dependence on temperature $T$ play a central role in
these materials as the thermopower is very sensitive to electron
localization. The search for new thermoelectric materials with large
$S(T)$ has motivated numerous studies on transition-metal oxides.
Recently the studies of oxides with high thermoelectric performance are
quite intense --- they led to the discoveries of the unusual combination
of high thermopower and high metallic carrier concentration in layered
cobaltites Na$_x$CoO$_2$ \cite{Ter97}, electron-doped titanates
\cite{Oku01}, and manganites \cite{Mai98}. It has been realized that
their large $S(T)$ is a consequence of narrow electronic bands which
motivated focusing on materials with high density of states (DOS) at
the Fermi energy or singularities at band edge due to their
low-dimensional electronic structure.

The DOS with a step at the band edge is found, {\it inter alia\/},
in transition-metal oxides due to the two-dimensional (2D) nature of
the electronic structure for either $t_{2g}$ \cite{Pav05} or $e_g$
\cite{Fei05} orbitals, or in frustrated 2D
lattices, such as triangular or kagome lattice. While the former might
play a role in future thermoelectric materials, the latter is the case
of the currently studied delafossites, such as YCuO$_2$ \cite{Sin08},
CuCr$_{1-x}$Mg$_x$O$_2$ \cite{Massc} and CuRh$_{1-x}$Mg$_x$O$_2$
\cite{Mai09}. These and some other oxides exhibit a $T$-linear
thermopower in agreement with the Fermi liquid (FL) theory, but
surprising systematic deviations from this behaviour have been observed
and their microscopic reasons are puzzling.
As reported by Kremer and Fr\'esard in their recent paper \cite{Kre12},
the FL behaviour is modified in the delafossites in the intermediate
temperature range, where the linearly (quadratically) extrapolated
thermopower (resistivity) exhibits a change in the slope between the
low and high temperature regime, suggesting a state which resembles the
FL. The authors call this state an {\it apparent Fermi liquid} (AFL)
\cite{Kre12}, as it occurs above the Fermi temperature. As found in
CuRh$_{0.9}$Mg$_{0.1}$O$_2$ \cite{Mai09}, a material in the AFL regime
exhibits a $T$-linear increase of the thermopower $S(T)$ but with a
finite offset at $T=0$, while the resistivity increases proportionally
to $T^2$, but again with a finite offset at $T=0$. This remarkable
behaviour makes the delafossites not only promising materials for
thermoelectric applications, but opens a new class of behavior that can
be observed at high or intermediate temperature.

The origin of this interesting novel behavior is investigated by
Kremer and Fr\'esard \cite{Kre12} using the temperature independent
correlation functions ratio (TICR) approximation to the thermopower
given by the Kubo formula \cite{Fre02}. It enables to evaluate this
quantity within the phenomenological approach, in contrast to advanced
studies based on the electonic structure found using density functional
theory \cite{Sin08}, or sophisticated approaches starting from model
Hamiltonians and using dynamical mean field theory. An advantage of the
phenomenological approach, introduced by Kremer and Fr\'esard
\cite{Kre12} to understand the origin of the AFL in weakly doped
materials, is that it allows one to extract the part
of the thermopower that follows from thermodynamics and to relate
it to the chemical potential even in very complex situations.

The phenomenological approach \cite{Kre12} is based on the generic form
of a quasi-2D model DOS $\rho(\omega)$ for frustrated lattices,
\begin{equation}
\rho(\omega)\simeq\frac{1}{W}\,\left(\rho_0+\rho_1\omega\right)\Theta(\omega)\,,
\label{dos}
\end{equation}
which is dominated by a discontinuity $\rho_0$ at the band edge
(at $\omega=0$), and increases linearly from it. A constant $W=1.0$ eV
is of the order of the bandwidth and is introduced here for convenience.
The band of a correlated insulator is gradually filled by doping $x$,
and
\begin{equation}
y=x/\rho_0
\label{y}
\end{equation}
is a dimensionless parameter in this theory. As expected, when doping
is large ($y>10^{-1}$), the discontinuity at the band edge has no
consequences and the usual FL behavior is obeyed that follows from the
Sommerfeld expansion. In contrast, for sufficiently small doping
$(y\simeq 10^{-2})$, the Fermi energy is close to the band edge,
leading to the AFL behavior. This regime broadens for increasing slope,
\begin{equation}
r_1=\frac{\rho_1}{\rho_0}\,W\,,
\label{r1}
\end{equation}
of the DOS, see Fig. 1
--- then the Fermi energy stays close to the band edge discontinuity in
a broad range of doping. Carrying out a thorough study of such a model
DOS by a Taylor expansion, the authors demonstrate \cite{Kre12}
that the $T$-linear thermopower can indeed be obtained either from a
high-temperature expansion (Boltzmann regime) or in a genuine
intermediate temperature range that requires a new approximation scheme.
They call it an {\it approximation of the polylogarithm difference\/}
(APLD), the name motivated by their expansion of doping $y$ using
polylogarithm functions.

%%%%%%%%%%%%%%%%%%%%%%%%%%%%%%%%%%%%%%%%%%%%%%%%%%%%%%%%%%%%
%%                      fig. 1
%%%%%%%%%%%%%%%%%%%%%%%%%%%%%%%%%%%%%%%%%%%%%%%%%%%%%%%%%%%%
\begin{figure}[t!]
\begin{center}
\includegraphics[width=8cm]{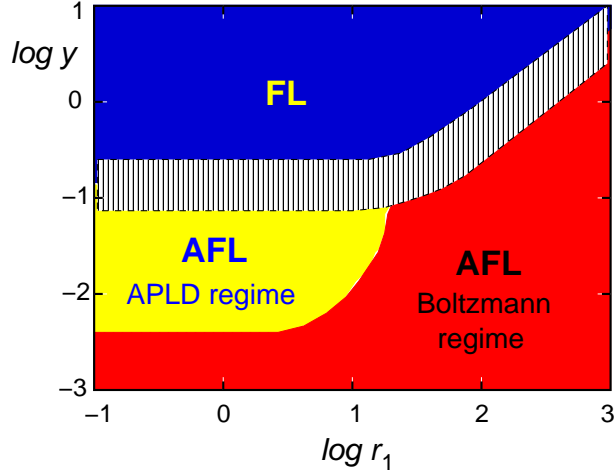}
\end{center}
\caption{(online colour at: www.ann-phys.org)
Artist's view of the phase diagram obtained by Kremer and Fr\'esard
\cite{Kre12}, with the Fermi liquid (FL) behavior at high doping $y$,
see Eq. (\ref{y}), while the apparent Fermi liquid behavior (AFL)
arises instead at small doping $y$, and may occur either in the range
of intermediate temperature (APLD regime), or at high temperature,
described correctly by the Boltzmann approximation (Boltzmann regime),
depending on the slope of the DOS near the band edge $r_1$, see Eq.
(\ref{r1}). Shadded area indicates the crossover regime between the FL
and AFL behavior. Note a double logarithmic scale of the phase diagram.}
\label{fig:aft}
\end{figure}
%%%%%%%%%%%%%%%%%%%%%%%%%%%%%%%%%%%%%%%%%%%%%%%%%%%%%%%%%%%%%%%

In spite of its conceptual simplicity and rather abstract
mathematical sophistication, the introduced concept of the AFL
represents an important advance in the theoretical description of
transport properties of real thermoelectric materials. Having put their
calculations to experimental test the authors obtain \cite{Kre12} that
Mg-doped delafossite CuRh$_{1-x}$Mg$_x$O$_2$ is in the AFL regime
described by the APLD, while Mg-doped delafossite
CuCr$_{1-x}$Mg$_x$O$_2$ falls instead in the AFL regime described by
the Boltzmann approximation. These two distinct regimes of the AFL are
shown in Fig. 1. The data of the thermopower of these two materials
show remarkably good agreement with the presented theory \cite{Kre12}
based on the DOS described by only two parameters defining the
phase diagram of Fig. 1: $r_1$ and $y$.

Regarding the specific heat a strong temperature dependence of the
linear coefficient $\gamma$ is also predicted by the theory due to
the onset of the AFL behavior under increasing temperature. It is
now challenging to verify the theory prediction that the breakdown
temperature for the FL theory increases with increasing doping, and
this could be tested for instance in CuRh$_{1-x}$Mg$_x$O$_2$.

The present discovery of the origin of the apparent Fermi liquid
behavior in hole-doped delafossites by Kremer and Fr\'esard
\cite{Kre12} adds a fundamental new concept to the search for
optimal thermoelectric materials. These compounds may be classified
in terms of an elegant and suggestive phase diagram with distinct usual
Fermi liquid and apparent Fermi liquid regimes
--- this classification and its relation to the electronic density of
states may also help in future search for new functional materials.

The author acknowledges support by the Polish Ministry of Science
under Project No. N202 069639.

%%%%%%%%%%%%%%%%%%%%%%%%%%%%%%%%%%%%%%%%%%%%%%%%%%%%%%%%%%%%
%%                    References
%%%%%%%%%%%%%%%%%%%%%%%%%%%%%%%%%%%%%%%%%%%%%%%%%%%%%%%%%%%%

\end{document}